\documentclass[]{spie}  

\usepackage[]{graphicx}

\title{Hard X--ray broad band Laue lenses (80--600 keV): building methods and
performances} 
\author{E.~Virgilli\supit{a}, F.~Frontera\supit{a}, P.~Rosati\supit{a}, V.~Liccardo\supit{b}, S.~Squerzanti\supit{c}, V.~Carassiti\supit{c}
, E.~Caroli\supit{d}, N.~Auricchio\supit{d} and J.B.~Stephen\supit{d} 
\skiplinehalf
\supit{a} \small\textit{Physics and Earth Sciences Department, University of Ferrara - Italy};\\
\supit{b} \small\textit{ITA-Instituto Tecnol\'ogico de Aeron\'autica, S\~ao
Jos\'e dos Campos - SP - Brasil};\\
\supit{c} \small\textit{ INFN - Istituto Nazionale di Fisica Nucleare - Sezione di Ferrara};\\
\supit{d} \small\textit{INAF, IASF Bologna - Italy}.
}

\authorinfo{E-mail to: virgilli@fe.infn.it}

\begin{document} 
\maketitle

\begin{abstract}
We present the status of the {\sc laue} project devoted to develop a technology for building a 20 meter long
focal length Laue lens for hard X--/soft gamma--ray astronomy (80--600 keV). The Laue lens is composed of bent crystals 
of Gallium Arsenide  (GaAs, 220) and Germanium (Ge, 111), and, for the first time, the focusing property of bent crystals 
has been exploited for this field of applications. We show 
the preliminary results concerning the adhesive employed to fix the crystal tiles over the lens support, the 
 positioning accuracy obtained and possible further improvements. The Laue lens petal that will be completed in a few 
months has a pass band of 80--300 keV and is a fraction of an entire Laue lens capable of focusing  
X-rays up to 600 keV, possibly extendable down to $\sim$20--30 keV with suitable
low absorption crystal materials and focal length.
The final goal is to develop a focusing optics that can improve the sensitivity over current  telescopes 
in this energy band by 2 orders of magnitude.
\end{abstract}

\keywords{Laue lenses, focusing telescopes, gamma--rays, astrophysics.}

\section{INTRODUCTION}
\label{sec:intro}

Past and present missions like $Beppo$SAX, XTE and INTEGRAL have demonstrated the key importance of 
 X--ray broad band (0.1--200~keV and beyond) observations to investigate and to explain a large number of 
astrophysical phenomena occurring in compact stars, active galactic nuclei or in diffuse emission
 (for an extensive list of astrophysical issues that are expected to be solved with focusing telescopes see 
e.g.~[\citenum{frontera10}]).
Nevertheless, these missions exhibited the limits of non focusing telescopes
in which, for instance, to increase the sensitivity by only a factor $\sim$3, the collecting area 
must be increased tenfold, with a consequent increase in terms of payload weight and, more importantly, 
of the detector noise. Focusing telescopes overcome this limitation, thanks to the physical separation 
between collecting area and sensitive surface. As the detector noise is roughly proportional to 
its volume, it turns out that focusing telescopes represent a key tool with which to maximize the signal-to-noise 
ratio. It has been shown that focusing telescopes in the 60--600 keV energy band could
overcome the sensitivity 
limits of the current generation of non focusing gamma--ray telescopes~\cite{frontera97, ubertini03}
by a factor $\sim$10--100.

Focusing telescopes in the soft X-ray energy band ($\sim$0.2--10 keV) are still operating since the late 90's. 
The upper limit has been extended to $\sim$70--80 keV with the employment of multilayer coatings 
(NuSTAR \cite{harrison13}).  Unfortunately, above 80 keV there is not yet a solid technology
comparable with the well established low and medium energy instrumentation.
For hard X-rays ($>$100 keV, but also applicable down to the 20/30--100 keV energy range) one concrete possibility to 
focus the radiation is offered by the use of Laue lenses which exploit Bragg diffraction by crystals. Besides order
of magnitudes increase in sensitivity, Laue lenses have the great advantage of 
improving the angular resolution of current X and gamma-ray telescopes, 
the best  now being obtained with coded mask telescopes (about 15 arcmin 
in the case of INTEGRAL/ISGRI).


\section{The LAUE project: an overview}
\label{sec:laue}

The {\sc laue} project\cite{Virgilli11b,Frontera13}, funded by the Italian Space Agency, is 
currently running in the {\sc larix} laboratory of the Physics and Earth Sciences Department of  Ferrara University (Italy). The main objective of the project is to develop a technology for building a Laue lens with a broad
energy band (70/100--600 keV) and long focal length (20 m), for astrophysics observations. 
In the project, the Laue lens is assumed to be made of a number 
of sectors, called petals. Assembling one of them, capable of focusing the radiation in the range
90--300 keV, is the starting goal of our project.
Within this activity, other main tasks have been conducted:   

\begin{itemize}
\item to realize a technology to produce a large number of diffractive crystals with high efficiency;
\item to find a method for assembling the crystals and  fixing them in place with high accuracy 
 in a relatively short time.
\end{itemize}


Flat crystals have the advantage of being relatively easily produced in large quantities with 
good reproducibility in terms of dimensions and mosaic spread. Unfortunately, their maximum reflectivity 
is limited to  50$\%$ \cite{Zachariasen, Malgrange02}~. Furthermore, a flat mosaic crystal tile does not display a focusing 
effect. Its diffracted image in the focal plane
 mainly depends on the crystal size and on the crystal mosaic spread which results in a defocusing 
 effect~\cite{Lund2005}~. In the {\sc laue} project, for the first time a large number of bent crystal tiles are being employed
in order  to exploit their focusing effect. Bent crystals can overcome the limit of flat crystals given that the diffracted 
 image  exhibited by a crystal with curved crystalline planes is smaller than the crystal cross section itself. 
 
 Bent crystals can be produced in a variety of methods~\cite{Smither05}~.  For our purposes,  mechanically bent crystals appear to be the most suitable. Self-standing Si and Ge perfect bent crystals were produced and characterized at the Sensor and  Semiconductor Laboratory ({\sc lss}, Ferrara - Italy) through mechanical grooving of one of
their surfaces~\cite{Bellucci11}~. An extremely uniform and highly reproducible curvature 
 is achieved  even if the grooves cause damage and  fragility of the crystals~\cite{Camattari14}~.  Instead, a self-standing curvature is obtained by {\sc imem/cnr}  Parma by means of a controlled  mechanical damaging of one surface of the sample~\cite{Buffagni11}~. This lapping procedure introduces defects in a
  superficial layer of a few microns, providing a highly compressive strain 
responsible for the convexity appearing in the crystal.

In the {\sc laue} project a flat frame is used as a support for the crystal tiles composing the single petal.
During the entire building phase the petal frame is kept fixed while a pencil beam of radiation is moved in front 
of the petal frame to mimic the presence of an astronomical gamma-ray source. The source of radiation is also 
directly used to control the correct positioning of the crystals. The total number of crystals that composes 
the designed  petal is 275 and the tiles are distributed in 18 concentric ring sections.

Positioning each crystal with the correct diffraction angle and keeping this orientation unchanged 
is the most challenging phase. Previous attempt to build Laue lenses have been made by using 
mechanical micro adjusters for positioning each tile with good results\cite{Halloin04}~. Nevertheless, 
using a micro-controller for each crystal appears to be a very hard solution for an entire Laue lens 
made of thousand  of crystals, considerably increasing the lens complexity and the total weight.
Therefore the strategy of the {\sc laue} project is to use a structural adhesive which works as an interface
between the frame support and the crystal itself.
The adhesive must have a fast curing time and as low a
shrinkage as possible, to minimize the building time and maximize the 
gluing accuracy, which is the most 
critical part. The elements that influence the gluing process are various. The temperature and the humidity of the 
building environment play a crucial role in the stability of the crystal/adhesive/support system. 
For this reason in the {\sc larix} facility a clean room with humidity and temperature stability control 
was set up (20 $\deg$ $\pm$ 1 $\deg$, relative humidity $\Phi$ = 50 $\pm$ 5$\%$).

\section{Configuration of the petal}

In Table~\ref{tab:configuration} the main properties of the petal  which is being built are reported. 
The energy pass band defines the inner and the outer radius of the  Laue lens petal. 
The crystal cross section has been chosen to be 30$\times$10 mm$^{2}$, with the longer side 
radially placed on the lens frame. The rectangular shape of the crystals gives a twofold advantage. Firstly, as
the focusing effect occurs only in the radial direction a shorter tangential dimension provides
a smaller defocusing factor in the latter dimension.
Secondly, a bigger radial dimension reduces the total number of crystals required, thereby reducing both the lens 
mounting time and the error budget potentially caused by each crystal misalignment contribution.
The designed lens focal length was set to 20 m. Consequently,  the crystal elements 
must have a 40 m curvature radius. The total number of crystals was equally subdivided into tiles 
made of Gallium Arsenide (220) (provided by {\sc imem/cnr}  Parma) and Germanium (111) (provided by 
{\sc lss} Ferrara). Employing the two types of crystals  allows a Laue lens to be built which is sensitive to the  
required energy pass band whilst keeping both inner and outer lens radius sufficiently small.  
Both {\sc imem/cnr} and {\sc lss} can provide 
the crystals with an accuracy in the curvature radius estimated to be within 
5$\%$ (40 $\pm$ 2 m).

\begin{table}[!h]
\caption{\footnotesize Main properties of the petal that is being built in the framework of the {\sc laue} project.}
\begin{center}
\begin{tabular}{|l|l|}
\hline
Materials  and selected planes          &   Ge(111), GaAs(220) \\
\hline
Energy passband                          &   80--300 keV \\
\hline
Focal length        	                 &   20 m\\
\hline
Petal inner/outer radius         	 &   20/80 cm\\
\hline
Petal dimension  (lens diameter)         &  $\sim$60 cm ($\sim$150 cm)\\
\hline
Crystal cross section                    &   30$\times$10 mm$^2$ \\
\hline
N$^o$ of crystals per petal (entire lens)    &   274 (5480)\\
\hline
N$^o$ of rings                           &   18 \\
\hline
Weight of the petal (entire lens)        &  1.3 kg (27.2 kg)\\
\hline
\end{tabular}
\end{center}
\label{tab:configuration}
\end{table}

\subsection{Alignment method} 
\label{alignment}
%
%
In our  reference system the $x$ axis is directed parallel to the x-ray beam impinging on the crystal 
while the $y$ and $z$ directions correspond to the  crystal principal axis, the former along the crystal 
longer dimension which is also the focusing direction.

\begin{figure}[!h]
   \begin{center}
   \includegraphics[scale=0.55]{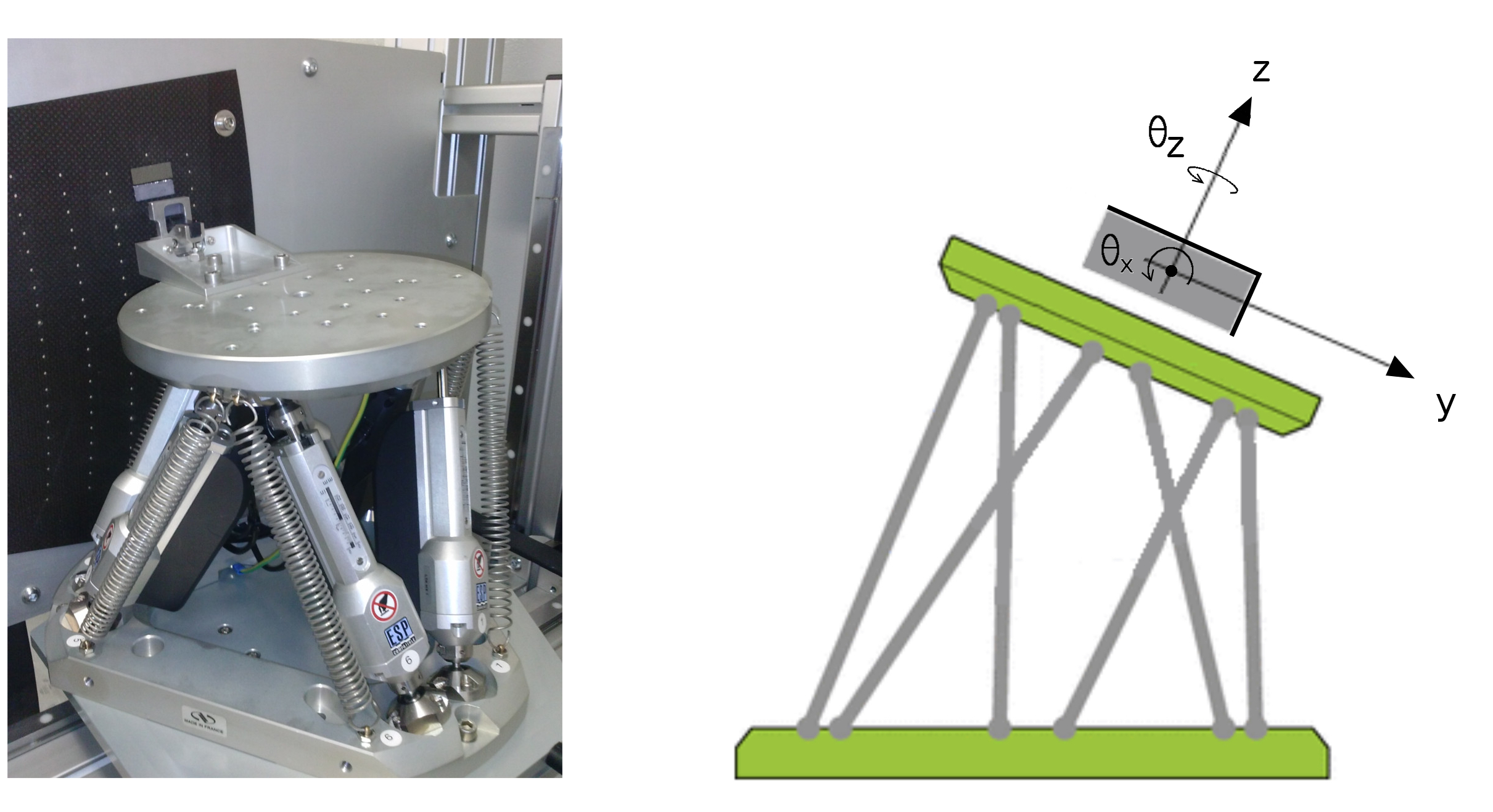}
     \caption{\footnotesize{Picture of the mechanical system (hexapod) adopted for 
     positioning the crystal over the frame. On its top, a crystal holder with a clamped tile 
 is mounted.}}
     \label{exapod}
\end{center}
\end{figure}

The goal of the {\sc laue} project is to develop a focusing lens capable of producing a photon distribution on the focal plane
with a spatial extension of $\sim$2~mm {\sc fwhm}, 
which corresponds to a  {\sc psf}  of 30 arcsec.  This value  includes  the crystal mosaicity
(15 arcsec for GaAs) and  all the contributions caused by the uncertainties during the alignment phases. 
All of the contributions must be considered separately and  minimized. The lens assembly method  
consists in the  positioning of each crystal tile on the lens frame
under the control of a gamma--ray beam.
%
%
The source employed must mimic
a source at infinity. For this reason the X-ray tube and the final collimator are
moved together in the $y$-$z$ plane in front of each crystal slot. Thanks to the mechanical precision of the carriages, the X-ray beam is parallel to the lens axis within about 1.5 arcsec.
Each single lens crystal is individually 
irradiated and appropriately oriented by using a hexapod (see Fig.~\ref{exapod}) in order to focus 
the radiation onto the correct point. The degrees of freedom of the hexapod  allowthe crystal  to be aligned 
with an accuracy of 0.01~mm for the three translations  and with $\sim$ 1.5 arcsec for the rotations. 
The spatial distribution of the diffracted signal and its correct positioning is observed with a 
1024$\times$1024 pixels X-ray imaging detector with 200$\times$200 $\mu$m$^2$ spatial resolution.

The most critical movement of the crystal is its rotation around its $z$ axis ($\theta_z$). A variation of $\theta_z$
changes the Bragg angle (and consequently also the diffracted energy) and results in a shift
of the diffracted signal on the detector plane which is directed along the $y$ axis. The 
measured shift  is proportional to the distance between the crystal and the detector itself.  
Instead, a variation of $\theta_x$  shifts the diffracted beam along a circle centered at 
the focal axis whose amount is proportional to the distance between the crystal and the axis of the 
lens.  The adhesives used to set the crystals suffer from a shrinkage effect that occurs 
during the curing phase and cause a change on both 
 $\theta_z$  and $\theta_x$ angles.

\begin{figure}[!h]
   \begin{center}
   \includegraphics[scale=0.3]{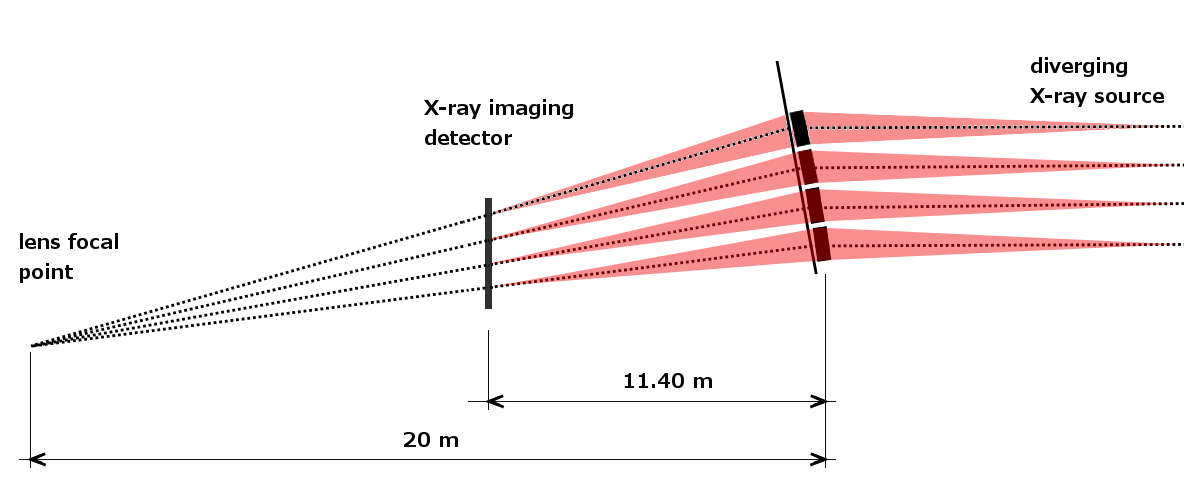}
     \caption{\footnotesize{Sketch of the Laue lens explaining the effect of focusing x-rays from a diverging 
     source. The collective behavior of the crystals is to focus the radiation at the distance of 20 meters from the lens but each crystal has its best focusing effect at a closer distance, depending on the distance between 
     the source of radiation and the crystal.}}
     \label{divergence}
\end{center}
\end{figure}

The procedure of mounting each crystal at the proper position is driven by the focusing power 
of the tiles. The best focusing is obtained at the nominal focal distance, if the source 
radiation is not divergent. Otherwise, the best focusing depends on the source divergence.
The focusing effect of bent crystals has been 
discussed and experimentally confirmed elsewhere~\cite{Virgilli15}
finding that the best focal distance depends on the combination of the sample  
curvature and the distance of the crystal from the radiation source.
This distance is  less than the nominal focal distance  and for our facility set-up  it turns out to be 
11.40 m (see Fig.~\ref{divergence}). 
Therefore, the mounting strategy for an astronomical Laue lens with 20 m focal length  
is to acquire the diffracted radiation from each crystal which are correctly oriented 
with the imaging detector positioned at  the distance of 11.40 m from the crystals. 
Using crystals with the cross section 30$\times$10 mm$^{2}$ the focusing effect makes 
the radial dimension  of the diffracted beam of the order of 1.3--1.5 mm (depending 
on the intrinsic mosaicity/quasi-mosaicity~\cite{ivanov05}) while in the tangential 
direction (where no focusing effect is expected) it reproduces the crystal dimension itself enlarged by the divergence effect.
%
%

Using the X-ray imaging detector and by fitting the diffracted profile  with a Gaussian function, the
position of the  barycenter can be estimated with an uncertainty of $\sim$0.5 arcseconds. On the basis 
of the previous considerations and considering the desired {\sc psf} dimension,
the barycenter of each crystal must be aligned at its proper reference pixel with 
an uncertainty of $\pm$2 pixels.

\subsection{Preliminary study of adhesives and methods} 

Various different methods to set the crystals at the correct diffraction angle have been considered. 
All the procedures were based on setting the crystals over a frame and 
the match between crystal and substrate was made with a structural adhesive.
As already pointed out in Sec.~\ref{alignment} the most critical effect while positioning 
a crystal is the rotation around the $z$ axis. Once the crystal is positioned, the curing process induces a 
differential adhesive shrinkage that results in a non negligible tilt around that axis.
In Table \ref{tab:glue} are reported the properties of the adhesives that have been tested.

\begin{table}[!h]
\caption{\footnotesize Properties of the tested adhesives for fixing the crystals to the petal frame.}
\begin{center}
\begin{tabular}{llll}
\hline
Adhesive name & properties   & curing time  & shrinkage \\
\hline
DEVCON Epoxy Gel  &  {\footnotesize Two-component}                                &  1 min      &   1-2 $\%$\\
DELO Automix 03 rapid thix  &  {\footnotesize Two-component}                       &  5 min      &   2 $\%$\\
Polyuretanic PUR 105   & {\footnotesize Two-component rigid}                      &   5 min            &   $<$ 1$\%$\\
Polyuretanic PUE 205   & {\footnotesize Two-component semi-rigid}                &    10 min           &   $<$ 1$\%$\\
DYMAX OP 61 LS  & {\footnotesize UV curing }                                &  10-20 sec &   $<$ 0.06$\%$\\
DYMAX OP 67 LS  & {\footnotesize UV curing }                                &  10-20 sec &   $<$ 0.06$\%$\\
\hline
\end{tabular}
\end{center}
\label{tab:glue}
\end{table}

\begin{figure}[!h]
   \begin{center}
   \includegraphics[scale=0.35]{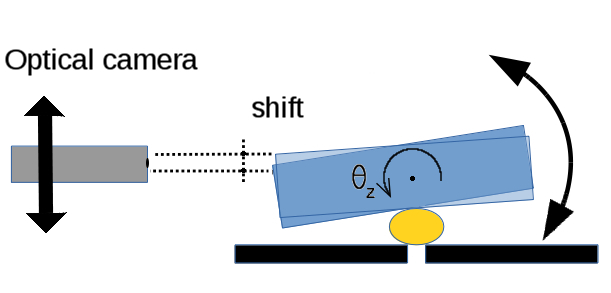}\\  
   \includegraphics[scale=0.055]{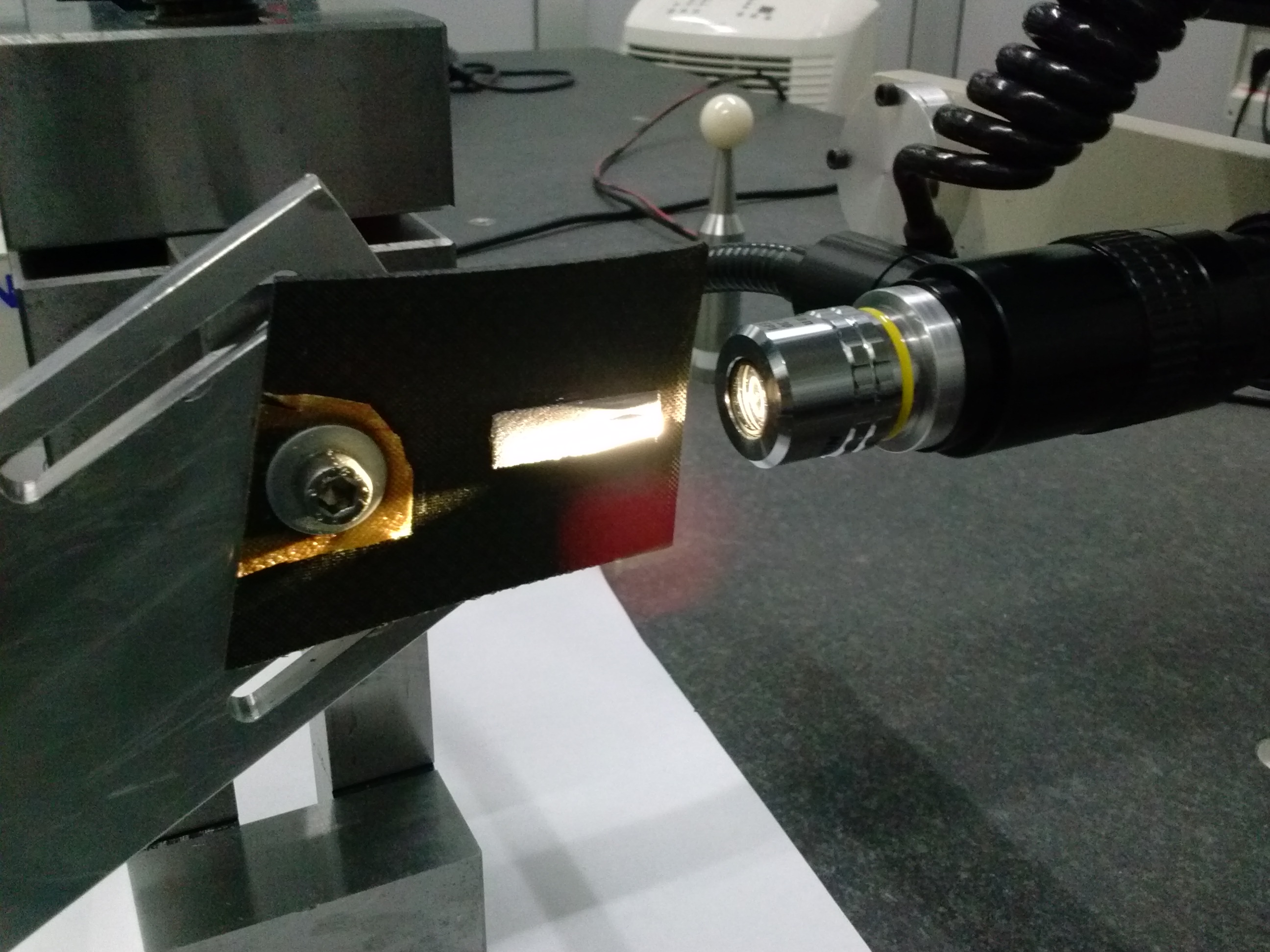}
    \includegraphics[scale=0.055]{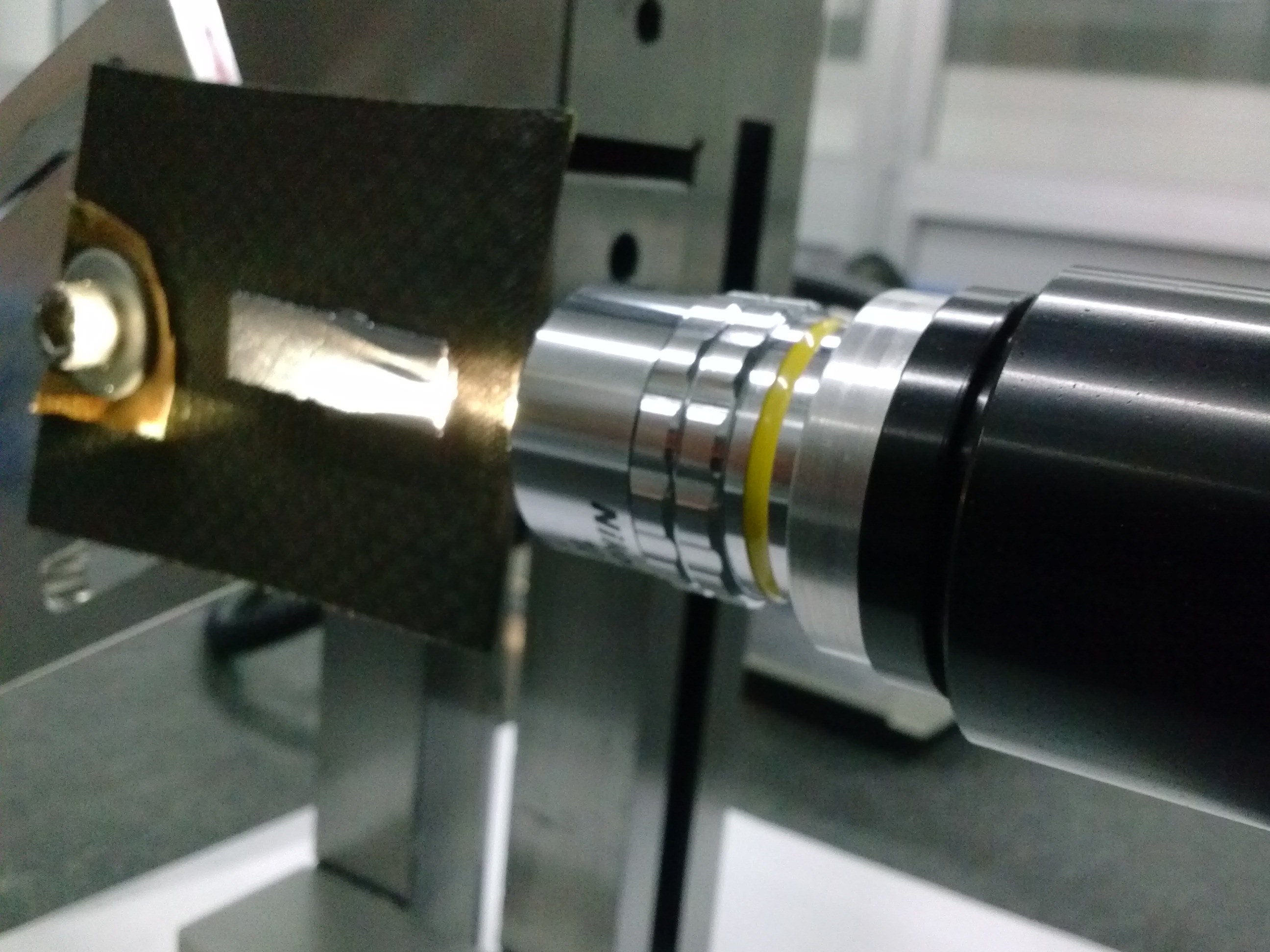}
     \caption{\footnotesize{Sketch and pictures of the optical camera 
     used to measure the linear shift of the crystal during the polymerization phase of the selected adhesive.}}
     \label{microscopio}
 \end{center}
 \end{figure}

Different substrates have been studied and compared, in order to minimize the discrepancy 
between the ideal positioning and the real outcome.
For each adhesive we analyzed pros and cons in terms of positioning accuracy, mounting 
time and stability with time.
The petal frame is made of 
carbon fiber 2 mm thick 
%
%
with a hole for each crystal. 
The fixing of the crystals to the frame was designed to 
be made by injecting the adhesive through the hole, with the crystal set at the correct position 
on the other side of the frame support.
At first, the structural DEVCON Epoxy Gel and the DELO Automix were used for their relatively short
curing time and the structural power. 
%
%
The stability of the gluing was initially performed by means of a high precision coordinate measuring machine  
 (Fig.~\ref{microscopio}).

 Each  crystal was positioned in such a way that the barycentre of the 
 diffracted image was in the center of the detector (pixel No. 512 in both directions). After 60 minutes 
 from the glue injection and at regular time intervals thereafter, 
%
%
the focused point was monitored with the  optical camera and, if needed,
the camera was repositioned and the shift recorded.

\begin{figure}[!h]
   \begin{center}
   \includegraphics[scale=0.3]{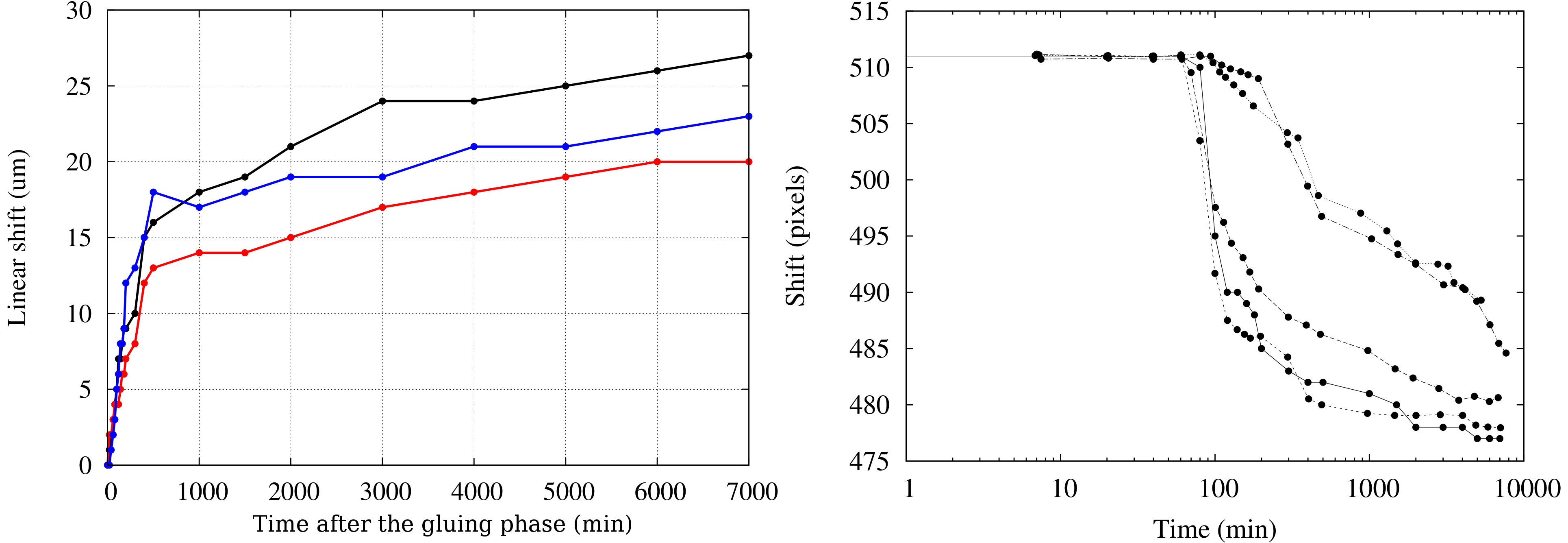}
     \caption{\footnotesize{{\it left}: The linear shift of a crystal measured with the optical camera, as a function of time.
     The total shift (20-25 $\mu$m) corresponds to a combination of rotation and translation of the crystal caused 
     by the glue shrinkage. {\it right}: Shift of the diffracted image (in detector pixels) measured with the 
X--ray beam 
     as a function of time for 5 different 
     crystals glued over a carbon fiber support. The glue injection was made through a hole performed through the support.}}
     \label{shiftottico_e_x}
 \end{center}
 \end{figure}

The results obtained with the DEVCON Epoxy Gel for 3 glued crystals is shown in the left panel of Fig.~\ref{shiftottico_e_x}. 
As can be seen,  
after $\sim$5 days the horizontal shift induced by the glue was roughly 20--25  $\mu$m. This amount would 
correspond to a shift of the diffracted image of $\sim$20 mm if the detector was positioned at the best focal 
distance of 11.40 m (misalignment of $\sim$5').
However, the measured shift 
can be due to a combination between a rotation $\Omega$ of the crystal, which is the main factor responsible of the diffracted beam movement, and a linear shift $\Lambda$ that 
does not significantly affect the position of the x-ray diffraction image. 
The cross check was done using the X-ray beam by gluing a sample of 5 crystals to a support frame, and testing the 
adhesive behaviour with the 
X--ray beam.

The results of the crystal monitoring  
is shown in the right panel of Fig.~\ref{shiftottico_e_x}. 
For the first 60 minutes the sample was set into the clamp and no motion was observed. 
After the release of the clamp, a common trend was observed which corresponded to a
decrease of the Bragg angle. Even though the trend is monotonically decreasing for all curves, the effect is not systematic.
The behavior is the same as suggested by the optical analysis but it confirms that the glue shrinkage behaves 
both in rotating the crystal and shifting it towards the frame support.
The gap between the reference pixel (512) and the experimental results
is in the range 30--40 pixels which corresponds to a shift of 6--8 mm ($\sim$100--140 arcsec), which is almost half 
of the amount estimated with the optical camera but is still far from our goal.

The carbon fiber was also used in a different configuration where the crystals were settled in   
cells and the glue was spread on the lateral edges of the crystals. 
In this configuration both the epoxy and  the polyuretanic adhesives were tested.
Thanks to the lateral gluing points the holding is more efficient even if the results are 
still within an uncertainty of 10--20 pixels (40-80"). The improved result with respect to the 
single gluing point can be explained by assuming that the single gluing point allows a pivoting of the
crystal around the fixing point while two lateral strips inhibit the shrinkage 
contributions.

Both the methods have shown the limits introduced by the glue shrinkage which is not negligible for our purposes.
For this reason we moved to a very low shrinkage single component adhesive which needs an activation given by  
UV light.

\section{Mock-up assembly}

\begin{figure}[!h]
\begin{center}
\includegraphics[scale=0.12]{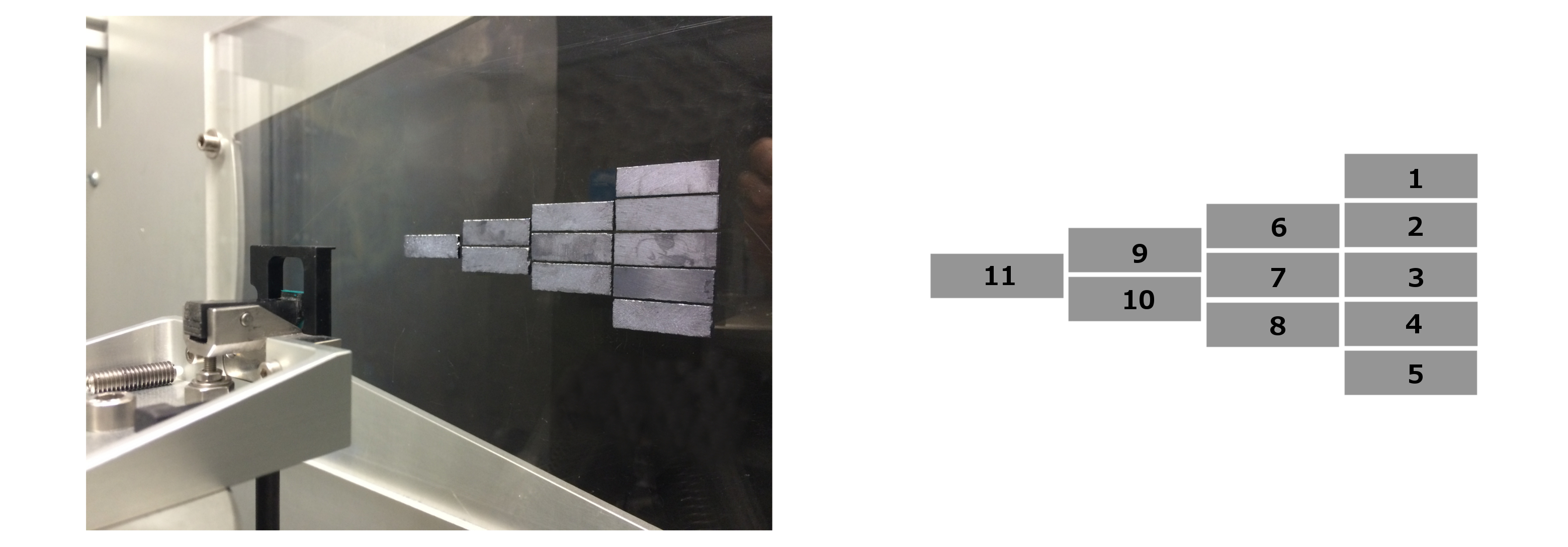}
\caption{\footnotesize{{\it Left}: picture of the mock-up made of 11 GaAs(220) bent crystals fixed on an 
     UV transparent PMMA flat frame. {\it Right}: drawing of the model with indicated the progressive crystal number reported with Fig.  \ref{misalignment}.}}
     \label{mockup}
 \end{center}
 \end{figure}

To minimize the shift and to quicken the process the adhesives DYMAX OP-61-LS and  DYMAX OP-67-LS  
have been tested. They are low-shrink, low-outgassing, low-CTE optical adhesives. Both are a single component paste and the curing 
occurs with optical and UV light.  
They have been chosen mainly for their extremely low shrinkage coefficient $<$ 0.08 $\%$ (other adhesives 
exhibited a shrinkage of 1--2$\%$). They have comparable physical properties but different viscosity (60000 cps 
for the OP-61-LS, 135000 cps for the OP-67-LS) and  curing time (from our experience the OP-67-LS 
cures $\sim$5 times faster than OP-61-LS).
The convenience of using a UV curing paste is that the adhesive can be applied to the crystal before the fine
alignment and the curing only occurs when the UV light is triggered.
As the adhesive reacts to UV light, a flat 10 mm thick transparent support made of 
polymethyl methacrylate (PMMA)is adopted. 
For the mock-up, 11 GaAs crystals were fixed over the frame with the configuration shown in Fig.~\ref{mockup} by using 
the hexapod system. For its higher viscosity and short polymerization time, the DYMAX OP-67-LS was selected.

Given that we were not interested in obtaining a tight packaging factor, the spacing between the 
tiles was set to 1 mm so as to avoid any problem of  accidentally  bumping one glued crystal while mounting the next.
As mentioned in Sec.~\ref{alignment}, during the mounting process each crystal was properly oriented in order 
to focus the radiation to the correct reference pixel. The crystal is first positioned
at 10 mm far from the frame for a rough alignment. The estimation of the diffraction angle allows the 
definition of the minimum distance at which the crystal can be set close to the frame, in order to minimize the 
adhesive thickness (typically between 50 and 100~$\mu$m). A fine alignment is made by 
measuring the position of the diffracted image barycenter with the crystal set at the final distance from the 
frame and correcting the diffraction angle with the hexapod until the correct  alignment is obtained.
A UV lamp with light guide is then irradiated at an appropriate  distance from the crystal, in order to slowly cure the adhesive while
minimizing the stress during this critical phase.

\begin{figure}[!h]
   \begin{center}
      \includegraphics[scale=0.3]{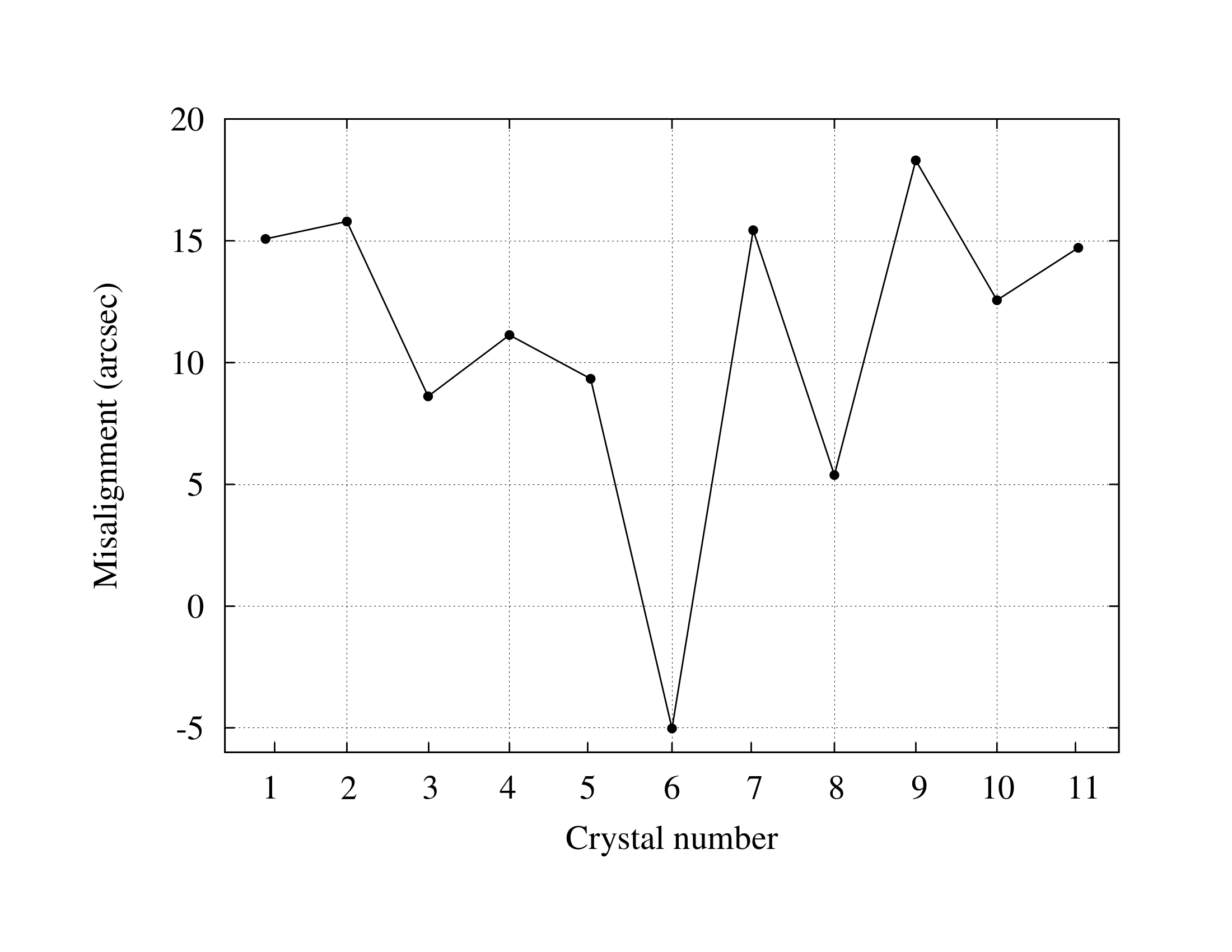}
     \caption{\footnotesize{Misalignment with respect to the ideal position for  the  11 GaAs crystals composing the  mock-up model after the procedure
     of gluing over the PMMA support.}}
     \label{misalignment}
 \end{center}
 \end{figure}

The crystal is then left inside the clamp for 30 minutes before being lightly released.
Thanks to the relatively fast curing time, this method allows a large number of crystals to be set per day (the mock-up model was 
built in 2 days) compared with other methods or adhesives that require 5-10 hours to fully polymerize.
The test of the correct positioning was done after $\sim$48 hours from the last glued tile and the results are 
shown in Fig.~\ref{misalignment} where the deviation between the target  and the measured position is expressed 
in arcseconds. It can be observed that all the crystals 
are correctly aligned within  $\sim$23 arcsec while more than 90\% of them is correctly aligned within 13 arcsec.

The results are still not satisfactory, but a significant improvement in the crystal gluing process 
has been obtained. Further work is needed.

%
%

\section{Conclusions}

In this paper we have described the  {\sc laue}  project which is devoted to finding and using a 
new technology for building Laue lenses for astrophysical observations. 
A dedicated facility was developed for the assembling and testing phase of Laue lenses. 
One of the most challenging tasks was to find a suitable crystalline material with a good efficiency 
at the energies of interest. Gallium Arsenide and Germanium with bent crystalline structure 
appeared to be valid candidates for their good efficiency with also the substantial feature of focusing capability. 
The possibility of focusing the radiation must be coupled with an assembling method 
whose main features have to be the high positioning accuracy and a convenient short mounting 
time  for each single crystal tile.
After a preliminary study of adhesives, we have found a satisfactory UV curable candidate to 
fulfill the requirement of a short polymerization time and obtaining a positioning accuracy within 15--20 arcseconds.  

Improvements in the mounting process chain are possible. One possibility is to use 
the carbon fiber frame with a large number of holes for each crystal. In this way the holes 
will ensure an effective curing phase of the UV adhesive and a minimal
radiation absorption. Using a different design for the holding system could be 
also helpful to firmly hold the tiles preserving their curvature.

Thanks to the experience gained in the past years and within the {\sc laue} project itself, 
we believe that Laue lenses will offer in the near future a new tool for
high-sensitivity astrophysical observations in the hard X--/soft gamma--ray energy
band, which will ultimately replace the current non-focusing telescopes.

\acknowledgments     

The authors wish to thank the Italian Space Agency for its support to the LAUE project.

\bibliography{references}
\bibliographystyle{ieeetr}


\end{document}